# L2AI: lightweight three-factor authentication and authorization in IOMT blockchain-based environment

Laleh Khajehzadeh, Hamid Barati, Ali Barati

Department of Computer Engineering, Dezful Branch, Islamic Azad University, Dezful, Iran

laleh.khajehzadeh@iaud.ac.ir, hbarati@iaud.ac.ir, abarati@iaud.ac.ir

A B S T R A C T

Medical Internet of Things (IoMT) is the next frontier in the digital revolution and is utilized in healthcare. In this context, IoT enables individuals to remotely manage their essential activities with minimal interaction. However, the limitations of network resources and the challenges of establishing a secure channel, as well as sharing and collecting sensitive information through an insecure public channel, pose security challenges for the medical IoT. This paper presents a lightweight multi-factor authentication and anonymous user authentication scheme to access real-time data in a blockchain-based environment. The scheme utilizes an insecure channel called L2AI. L2AI ensures security and efficiency while enhancing user anonymity through the use of pseudo-identity and dynamic indexing. The proposed method supports highly scalable systems with an efficient user registration process, allowing authenticated users to access both existing and newly added system entities without additional processes. Although the scheme is primarily designed for large systems, such as health infrastructure, it is also suitable for resource-constrained devices. The scheme relies on one-way cryptographic hashing functions and bitwise XOR operations. Additionally, a fuzzy mining algorithm is employed on the user side to verify the user's biometric information. L2AI adopts the "Real-Or-Random (ROR)" model for security proof and employs BAN logic for proof of authenticity. Formal security verification is conducted using the "Automatic Validation of Internet Security Protocols and Programs" (Proverif) tool, complemented by informal security analysis demonstrating the proper functionality of L2AI. Furthermore, a simulator implemented in Python is used to illustrate the practicality of the design. Ultimately, L2AI offers desired features and achieves mutual authentication with low computation and communication costs compared to existing schemes.

## 1. Introduction

Internet of Things (IoT) technology is a network of interconnected smart devices based on the Internet that offers tremendous benefits. IoT has already played a prominent role in modern daily life and established itself as a new paradigm. It is predicted that by the end of 2025, the number of connected IoT devices worldwide will exceed 30 billion [1-6].

Recently, IoT technologies, aiming to increase intelligence and efficiency in healthcare, have led to the development and emergence of new perspectives in this sector. The Internet of Medical Things (IoMT) comprises a large number of connected heterogeneous smart devices or wireless sensors and robots. These devices seamlessly collect and exchange sensitive data through the Internet in various fields to achieve specific goals. The collected data from IoT sensors is directed to servers for storage and further processing. Authorized users can then access the desired data from the relevant database [7-9].

However, despite the significant benefits of IoT, concerns regarding security and privacy violations arise, particularly when it is utilized in medical settings. Mutual authentication mechanisms are employed as the first line of defense to verify the identity of devices and users, ensuring the legitimacy of each device and privacy obligations [10-12].

Another crucial aspect and a reliable foundation for ensuring secure communication in IoT environments is the control of information exchange access. Authorization and access control mechanisms grant different access rights to users based on data importance and privacy sensitivity. These mechanisms verify whether a user has the appropriate permissions to access the data or not [13-14].

"Anonymity" is an essential measure to comply with the EU General Data Protection Regulation. This technique plays a vital role in maintaining data privacy and safeguarding sensitive data and user identities against passive attacks. Furthermore, "anonymity" strengthens the mutual authentication mechanism, making it more secure and preventing attackers from tracing the user and the service provider [15-17].

Blockchain, as a distributed ledger, serves as a storage medium for transaction-related records, ensuring accessibility and immutability. It has become an infrastructure in various fields due to its robust security mechanisms such as "immutability," "decentralization," and other important elements. In a blockchain, each "block" represents a transaction or record that must be stored in a database. The encryption technology employed in blockchain ensures the integrity of data, making it resistant to changes. The consensus mechanism, a fundamental technology in the blockchain, prevents failures in decentralized distributed systems through a consensus algorithm within the blockchain network [18-22].

### 1.1. Motivation

So far, several authentication schemes and key agreements have been proposed to address these requirements. However, most of these approaches focus only on verifying security and integrity during setup and registration through a "secure channel" and overlook the challenges associated with establishing and creating the secure channel itself. Creating a secure channel is a significant challenge in the context of IoT. A secure channel allows for remote access and management of medical activities in real-time with minimal human involvement. However, the collection and sharing of sensitive data among IoT servers, users, and sensors over a public insecure channel expose them to various security risks and threats. Medical data, in particular, must remain confidential due to its critical nature, as it directly impacts the health and lives of patients. Authentication serves as a vital mechanism to mitigate the aforementioned risks.

Additionally, it is important to consider the limitations of processing and computing power in IoT devices, as they have constrained resources. In real-time applications, these devices require minimal calculations for optimal performance. Therefore, the use of lightweight security operators such as bitwise XOR operations and cryptographic hash functions is highly desirable in such environments. However, many existing authentication schemes remain insecure and vulnerable to known malicious attacks. Hence, designing a lightweight and secure authentication mechanism for modern medical care applications is of utmost importance.

Our objective is to design a lightweight and secure three-factor mutual authentication and access authorization approach in an IoT-based environment to address the aforementioned challenges and facilitate modern medical care. This approach ensures user identity verification while preserving anonymity. Considering that the registration stage occurs over an insecure public channel, the proposed scheme incorporates necessary security features to guarantee session security and enhance the overall security level. Furthermore, the scheme is designed to optimize performance and carry out other essential operations, such as protecting user identity and basic information, ensuring smart card security, and improving overall system performance.

To determine the access level, users receive corresponding tokens from the server during the registration request. These tokens are issued based on users' roles and permissions within the healthcare system. The proposed scheme utilizes lightweight encryption techniques to protect privacy and authenticate users. This approach enhances the strength and resilience of the IoT network against security threats and attacks, even with limited resources. Additionally, a new session key is generated between legitimate users and servers to establish secure and real-time communications. This key ensures the confidentiality of previous communications and prevents their disclosure. The architecture of the proposed protocol is based on the blockchain platform, which offers higher security for data protection and enables decentralized and transparent management, including the creation of smart cards for each user. This blockchain-based architecture safeguards smart cards from unauthorized alterations or tampering.

### 1.2. Contributions

Our contributions can be summarized as follows:

1. We propose a lightweight three-factor authentication and authorization protocol called L2AI for real-time data access in an IoT-based environment on a blockchain platform. The protocol utilizes cryptographic hash functions, bitwise XOR operations, and a biometric hashing mechanism to incorporate the user's biometric information.

2. In L2AI, we employ a lightweight authorization mechanism to provide convenient access control, safeguarding sensitive system resources from unauthorized access. The user's pseudo-identity name and private key are dynamically updated during each session key agreement to ensure a one-time pad. Our scheme guarantees user anonymity and disconnection.

3. To enhance security, transparency, and scalability, we utilize a private blockchain to record information in smart cards and prevent information manipulation.

4. We have conducted a formal security analysis of L2AI using the Real-or-random (ROR) model and established the correctness of the proposed scheme using (BAN) logic. The protocol has also been tested using the Automated Validation of Internet Security Protocols and Applications Toolkit (AVISPA). An informal security analysis demonstrates L2AI's resilience against multiple attacks.

5. We have implemented L2AI extensively using Python and measured its impact on network performance parameters. Additionally, we have conducted a comprehensive comparison between our proposed protocol and other related protocols[...] in terms of computation cost, communication cost, security requirements, and resistance to attacks. The results indicate that our scheme is more efficient, secure, and practical.

### 1.3. Roadmap of the paper

The structure of our paper is organized as follows: Section 2 introduces the network and threat models, along with the proposed work authorization mechanism, L2AI. Section 3 provides basic information about L2AI. Section 4 offers a detailed explanation of the steps involved in L2AI. Section 5 presents both formal and informal security analyses. Section 6 compares the performance of L2AI with other related schemes. Section 7 discusses network analysis and simulation results using Python. Finally, Section 8 concludes the paper, summarizing the key findings and contributions.

### 1.4. Related work

The proposed method in [23] addresses the security issues present in Wu et al.'s scheme. It involves four phases: Initialization, Registration, Login and Authentication, and Password Change. In the Initialization phase, the gateway (GW) generates an elliptic curve group, secret keys, and hash functions. During the Registration phase, a user (Ui) provides their identity and password, and the GW issues a smart card. Specific values are stored for sensors (Sj). In the Login and Authentication phase, the user's smart card generates session keys and computes parameters, which are then sent to the GW for authentication. This method employs robust cryptographic techniques to prevent attacks, ensuring secure communication.

In [24], an ECC-based RFID authentication protocol is presented to achieve mutual authentication and address security vulnerabilities in existing protocols. This protocol fulfills various security requirements, including mutual authentication, scalability, forward security, data confidentiality, tag anonymity, data integrity, and availability. It demonstrates resilience against attacks like replay, de-synchronization, cloning, location tracking, server spoofing, tag masquerade, modification, and DoS attacks. The protocol's analysis shows reduced computational overhead, communication costs, and storage requirements, making it suitable for RFID tags with limited resources.

In [25], a secure two-factor authentication scheme is presented for IoT environments. The scheme utilizes hash-chain techniques and forward secure protocols to enhance user authentication security. It involves three steps: registration, authentication, and update. During the registration process, the user's device and gateway exchange information and store cryptographic values derived from the user's identity, hashed passwords, and random nonces. In the authentication phase, the user's device sends hashed and encrypted values derived from stored credentials and fresh random numbers to the gateway for verification. Successful authentication prompts both entities to update their stored values for forward security, preventing replay attacks and compromising future sessions.

In [26], a lightweight privacy-preserving scheme (L-PPS) for UAV networks is introduced, utilizing Chebyshev Chaotic Maps. The scheme aims to establish a secure and efficient authentication mechanism within Internet of Drones (IoD) environments. L-PPS operates in a smart IoD environment and involves a key distribution process for UAV systems, utilizing drone-nodes and mobile-sinks. It addresses security concerns by providing a valid authentication period, ensuring persistent authentication, reducing computation costs of symmetric encryption/decryption, and facilitating mutual authentication and session key agreement between communication entities. The scheme incorporates continuous authentication with a secret token to enhance transmission rates and prevent packet loss. The validity of the scheme is confirmed through Scyther verification and simulation analysis, demonstrating superior performance metrics such as packet delivery ratio and end-to-end delay.

In [27], a lightweight multi-factor authentication and authorization scheme called LMAAS-IoT is presented. It is specifically designed for enabling real-time data access in IoT cloud-based environments. The scheme ensures robust security, efficiency, and user anonymity through the utilization of dynamic indexing. It is capable of supporting highly scalable systems and incorporates various cryptographic hash functions, bitwise XOR operations, as well as a fuzzy extractor algorithm for biometric verification. The security analysis of the scheme includes the Real-Or-Random model, BAN-logic, and AVISPA tool. LMAAS-IoT has been implemented using the NS-3.31 simulator, successfully demonstrating its practicality. This scheme offers desirable attributes such as mutual authentication, while significantly reducing computation and communication costs when compared to existing alternatives.

In [28], a user authentication protocol is presented with the aim of enhancing security in coal mine environments using IoT-enabled wireless sensor networks (WSNs). The protocol addresses various security threats, including insider attacks, impersonation attacks, and device theft. It incorporates mutual authentication and key agreement processes to ensure secure communication between users and sensor nodes. The scheme has undergone formal verification using the ProVerif tool, and its performance has been compared to other protocols in terms of computational, storage, and communication costs. It has demonstrated higher reliability and robustness compared to existing schemes.

In [29], a hierarchical key management and authentication method is proposed for wireless sensor networks. The method divides the network into non-overlapping zones and establishes connections between network nodes using two keys within each zone and between zones. The zone key is shared between zone members and the zone manager, while the inter-zone key is shared among manager nodes for inter-zone communication. The method includes lightweight authentication operations between zone members and the zone manager for intra-zone communication. Its objectives are to enhance network security, reduce computational load and energy consumption, and improve scalability, flexibility, and efficiency in wireless sensor networks.

In [30], an authentication-based secure data aggregation method is presented for the Internet of Things (IoT). The method comprises three phases: constructing a star structure within each cluster with unique encryption keys, intra-cluster communication with encrypted data and key updates, and enhancing inter-cluster communications through an authentication protocol. The primary objectives of the method are to enhance network security, reduce energy consumption, and improve the network lifetime. It utilizes a hierarchical framework for data aggregation, incorporates the rail fence cipher algorithm to enhance security, and implements an efficient authentication method for communication between cluster heads.

In [31], an anonymous user authentication and key agreement scheme specifically tailored for the Industrial Internet of Things (IIoT) is introduced. This method enhances security through the utilization of hash functions, elliptic curve encryption, and fuzzy biometric extraction. By incorporating a pseudonym tuple database, the scheme enables dynamic user joining and ensures anonymity while effectively mitigating key loss and device capture attacks. The proposed method undergoes rigorous formal security analysis using the BAN logic and ROR models, confirming its robustness against existing threats. With a focus on high communication efficiency and functionality, this scheme is designed to cater to the unique requirements of IIoT environments.

In [32], an enhanced three-factor authentication protocol is specifically introduced for IoT environments. The protocol aims to address security vulnerabilities identified in Mirsaraei et al.'s scheme, including untraceability and key compromise impersonation resistance. The method leverages the elliptic curve cryptosystem (ECC) to enhance security features, such as perfect forward secrecy, while also minimizing computational and communication overhead. To ensure its robustness, the protocol undergoes formal verification using the ROR model and Proverif tool, demonstrating its resilience against various attacks and its efficiency in IoT devices with limited resources.

In [33], a Firmware-Secure Multi-Factor Authentication (FSMFA) protocol is proposed for IoT devices, which integrates PUF (physical unclonable function) and firmware integrity to enhance both physical and software security. This approach enables mutual authentication and key negotiation between devices and servers, effectively addressing threats such as cloning, tampering, and malware. The protocol's security and efficiency are substantiated through the use of BAN logic and ProVerif tools, demonstrating its superiority compared to existing protocols.

In [34], the LDA-2IoT (Level Dependent Authentication for IoT) scheme is introduced, which utilizes ECC-based two-factor authentication specifically designed for IoT environments. The scheme enables users at a particular hierarchy level to access sensors at or below their level, reducing the requirement for multiple registrations. It employs lightweight cryptographic operations to ensure secure data transmission and mutual authentication between IoT devices and users. Security analyses using the Dolev-Yao channel and the Real or Random model validate its robustness. Real-time implementation with the MQTT protocol demonstrates high efficiency in terms of network throughput, computation, and communication costs.

In [35], a secure mutual authentication and session key negotiation scheme is introduced for IoT-based Wireless Sensor Networks (WSNs), utilizing Elliptic Curve Cryptography (ECC). This method aims to address vulnerabilities present in existing protocols, including user impersonation, stolen smart card, database, and insider attacks. Through informal and formal security analyses, the proposed scheme demonstrates resilience against multiple threats. Furthermore, it showcases improved performance metrics when compared to contemporary methods.

In [36], ECC-based Authentication Scheme (ECCbAS) is introduced as a solution to address vulnerabilities found in a previously proposed protocol by Sureshkumar et al. for healthcare IoT systems. ECCbAS mitigates threats such as traceability, integrity contradiction, and de-synchronization attacks through a combination of formal and informal security analyses. The protocol utilizes two separate channels for data transmission: one during the enrollment phase and another for communications during authentication and password change phases. ECCbAS demonstrates significant improvements in both security and performance compared to its predecessor, effectively countering potential security attacks.

In [37], an enhanced lightweight and anonymity-preserving user authentication scheme is proposed specifically for IoT-based healthcare environments. This method builds upon Masud et al.'s protocol by integrating advanced security measures such as biometrics, fuzzy extractors, secret salts, and physically unclonable functions (PUFs) to address vulnerabilities like session key disclosure, offline password guessing, and traceability attacks. The protocol incorporates both heuristic and formal security analysis to ensure robustness and efficiency in resource-constrained scenarios. However, critical analysis reveals that while the protocol enhances security and maintains operational efficiency, it may still face challenges in large-scale implementations and against sophisticated physical attacks.

In [38], a secure and lightweight mutual authentication and key establishment protocol is introduced for cloud-assisted IoT systems. The protocol is designed to overcome the resource constraints of IoT devices and ensure secure communication over insecure channels. Key features of the protocol include device anonymity and untraceability, efficient password updates, and device revocation. It leverages elliptic curve cryptography (ECC) to achieve security objectives. The protocol's security is validated through informal analysis, BAN logic-based verification, and the AVISPA tool, demonstrating its resistance against various attacks. Performance analysis indicates improved efficiency when compared to existing protocols.

In [39], a Lightweight Authentication Protocol for IoT Devices (LAPE2D) is presented as a lightweight authentication protocol designed to secure IoT devices. The protocol utilizes lightweight cryptographic mechanisms, specifically one-way hashing and XOR operations. It supports various multi-device authentication scenarios, including end node to gateway, gateway to gateway, and node to node through the gateway, ensuring secure communication with session keys. The protocol's robustness is demonstrated through message exchange processes and cryptanalysis against common network intrusion attacks. It exhibits improved efficiency in terms of communication overhead and computing complexity compared to existing methods. However, it is important to note that while the method enhances security, it may still face challenges in highly dynamic and large-scale IoT environments.

In [40], an improved lightweight authentication protocol is designed for IoT-based medical care systems. The protocol utilizes a one-way hash function and XOR operations to achieve secure communication between medical practitioners and sensors through insecure public channels. The protocol involves mutual authentication between users and sensor nodes, password update mechanisms, and node revocation features. It aims to address the limitations of Masud et al.'s method, which lacked robust password management and was vulnerable to node clone and replication attacks. The proposed protocol enhances security by generating session keys post-authentication and incorporates mechanisms to resist various attacks.

**Table 1. Related works.**

| Ref | Advantage | Disadvantage |
|---|---|---|
| [23] | The method enhances security by preventing offline password guessing, user forgery, and gateway forgery attacks through advanced cryptographic measures. | The proposed scheme might increase computational overhead and complexity due to multiple cryptographic operations and processes involved. |
| [24] | Provides robust security against various attacks, Low computational overhead, Suitable for resource-constrained RFID tags | Implementation complexity, Potential challenges in large-scale deployment, Requires secure ECC implementation to maintain effectiveness. |
| [25] | Robust against replay and man-in-the-middle attacks, Ensures forward security, protecting future sessions, Suitable for resource-constrained environments like IoT. | Heavily relies on the security of the hash functions, Computational overhead may affect extremely limited devices, Requires proper synchronization, challenging in high-latency networks. |
| [26] | L-PPS offers enhanced security with low computational and communication costs, robust mutual authentication, and session key agreement, ensuring reliable and efficient performance in IoD environments. | L-PPS might face implementation challenges in highly dynamic or large-scale IoD networks, and its reliance on specific cryptographic methods could limit flexibility and adaptability in diverse operational contexts. |
| [27] | Strong security through chaotic maps, Resistance to various attacks (e.g., replay, insider, and stolen smart card attacks) | Implementation complexity, Potentially higher computational resource requirements |
| [28] | Provides robust security against various attacks, Formal verification using ProVerif enhances trustworthiness, Efficient in terms of computational, storage, and communication costs. | May require significant initial setup and configuration, The complexity of the protocol could lead to implementation challenges, Dependence on secure key management practices. |
| [29] | Enhanced security through hierarchical structure, efficient key management, and authentication. | Reliance on shared keys may pose risks in case of compromise, potential overhead in managing zone structures and maintaining key updates. |
| [30] | Improved energy consumption, end-to-end delay, flexibility, packet delivery rate, and number of alive nodes. | Additional overhead and complexity for key management and authentication, lack of scalability in large-scale IoT deployments |
| [31] | High computational and communication efficiency, supports user addition/revocation, and ensures complete anonymity and untraceability. | Potential complexity in implementation, reliance on secure initial setup, and vulnerability to sophisticated future attacks. |
| [32] | Provides perfect forward secrecy, reduces computational and communication overhead, and enhances security against known attacks. | Implementation complexity, potential higher initial setup costs, and reliance on ECC which may not be supported universally. |
| [33] | Enhances security with minimal overhead; robust against physical and software attacks; ensures lifecycle device security. | May require specialized hardware for PUF; potential complexity in integrating fuzzy extractors; might not suit all IoT scenarios. |
| [34] | Reduces access control complexity, lightweight due to ECC, efficient for large-scale IoT deployments. | Potential scalability issues with very large hierarchies, dependency on secure ECC implementations, initial setup requires careful configuration. |
| [35] | The proposed ECC-based scheme offers enhanced security, reduced computation, efficient communication, and lower storage overheads. | Despite its strengths, the scheme may face implementation complexity and potential increased initial setup costs due to ECC's computational requirements. |
| [36] | Enhanced security and performance, robust against traceability, integrity contradiction, and de-synchronization attacks. | Potentially increased complexity in implementation and maintenance, reliance on |
| | | the secure management of two communication channels. |
| [37] | Improves security and efficiency in IoT-based healthcare; utilizes advanced cryptographic techniques. | Potentially challenging in large-scale applications; may not fully withstand sophisticated physical attacks. |
| [38] | Enhanced security with device anonymity, untraceability, and efficient password updates; validated through formal and informal methods. | Dependent on ECC, which may still pose computational challenges for highly constrained devices; lacks practical implementation details and real-world deployment metrics. |
| [39] | Enhanced security and efficiency in communication overhead and computing complexity for IoT devices. | Potential challenges in scalability and adaptability in highly dynamic and large-scale IoT environments. |
| [40] | The protocol is lightweight, enhances security through robust authentication and session key generation, and includes password management and node revocation features, ensuring comprehensive protection. | It may require higher computational resources than simpler protocols, potentially impacting system performance. Implementation complexity and maintenance could also be higher due to additional security mechanisms. |

## 2. System models

In this section, we show the proposed network model, threat models, and authorization mechanism L2AI.

### 2.1. Network model

The L2AI comprises four entities: an authentication server, a smart gateway/mobile terminal, IoT sensors, and users (including patients and medical staff). Figure 1 illustrates the network model of L2AI.

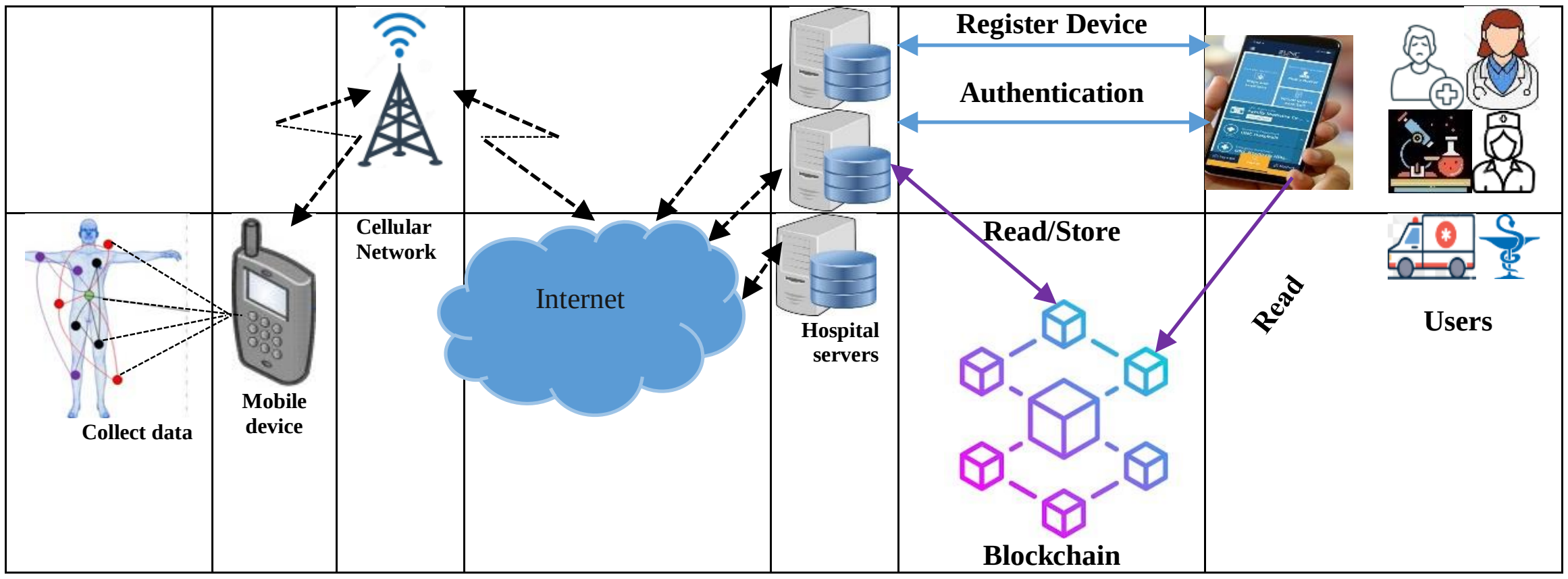

**Fig. 1. The network model of** L2AI.

The authentication server plays a crucial role in managing device and user information within the entire IOMT environment. To become a legitimate user, both hospital users and patients must first register themselves on a trusted hospital server. During the registration stage, the server generates the necessary credits (tokens) based on personal credit information and sends them to the user's smart gateway/mobile terminal, which is registered with the server. Additionally, the authentication server, in accordance with the preferred policy, manages user privileges (Ui) to provide an authorization mechanism alongside authentication, ensuring the security of sensitive system resources.

To initiate the registration process, the user connects to the authentication server through a smart gateway/mobile terminal, which can be an application or a web page. Mutual authentication between users is performed before any information exchange takes place. The network model in this design encompasses three types of communication: (1) communication between the user and the server, (2) communication between the server and IoT sensors, and (3) communication between the smart gateway/mobile terminal and the user. Secure communication requires key management between the user and the server. Once mutual authentication is established, both parties generate a new session key for secure communication. The server stores the required user data in a blockchain platform. Subsequently, the user can indirectly connect to IoT devices through the server and exchange data. This scenario enables hospital users (such as doctors and nurses) with different access levels to monitor patients' conditions.

To enhance security, the authentication protocol is presented in four steps: registration, login and authentication, updating biometrics and password, and updating the access level. Figure 2 provides a concise overview of the framework flow diagram for L2AI.

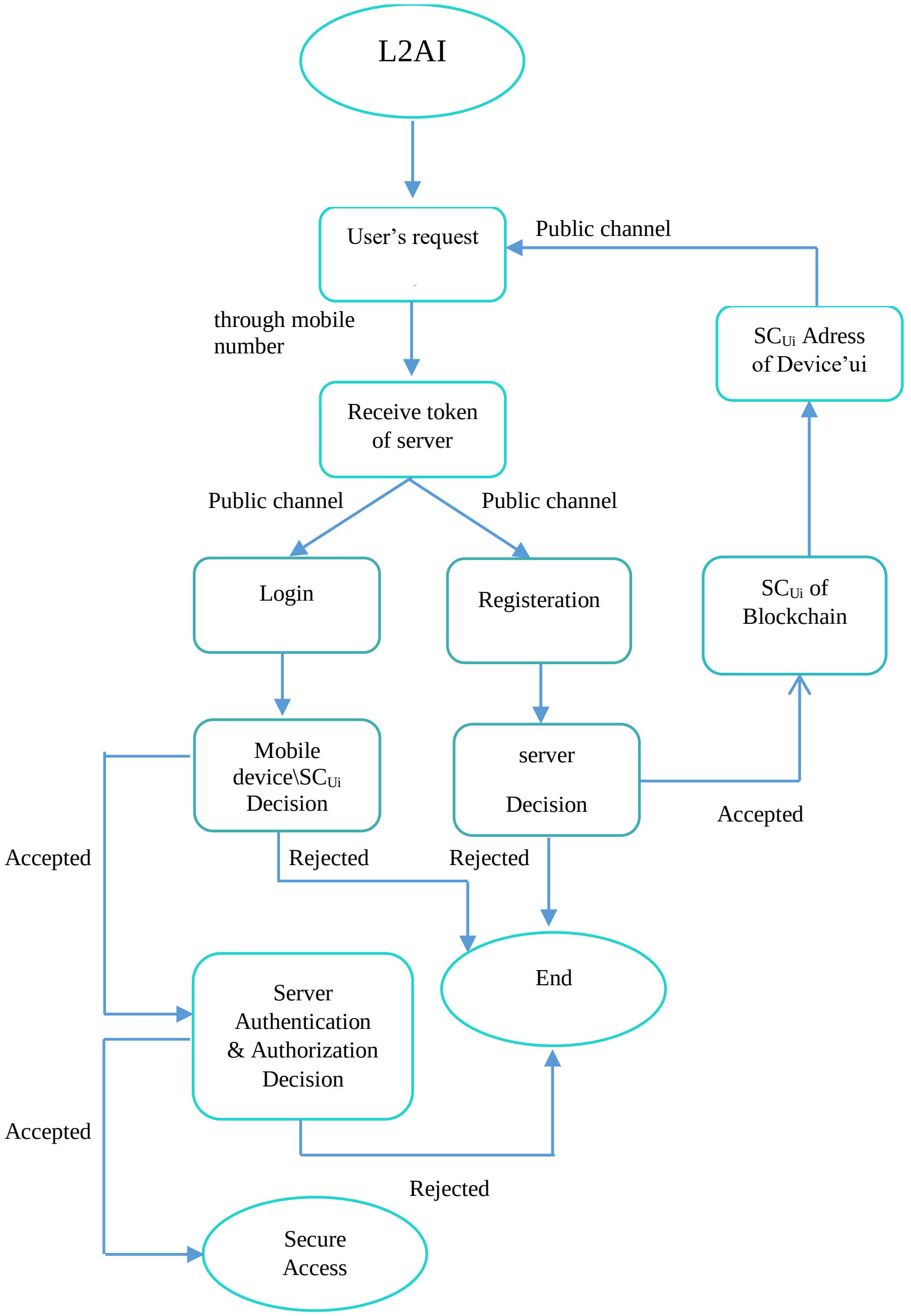

**Fig. 2. The flow diagram of L2AI framework.**

## 2.2. A Layer-based Overview

The architecture of the proposed system illustrated in Figure 3 is composed of six layers as follows:

The first layer is the physical layer, comprising wearable health devices. These devices monitor vital signs such as breathing rate, heart rate, and blood pressure in real-time, ensuring accurate data collection.

The second layer is the connection layer, which includes a healthcare IoT server and a smart gateway/mobile terminal. After successful authentication, the wearable health devices transmit the collected vital signs data to the healthcare server through the smart gateway/mobile terminal, establishing a reliable connection.

The third layer is the Users Layer, encompassing various individuals and entities, such as patients, doctors, pharmacists, laboratory technicians, and other medical institutions, all connected to the Internet of Medical

Objects server. In this layer, authentication operations take place between the users and the IoT server. Once authenticated and authorized, users can securely exchange data within the network.
The fourth layer is the Blockchain network layer, utilized to store user IDs, smart card information, and access level permissions in the proposed protocol. Blockchain technology addresses several security concerns in IoT. Its distributed network structure ensures the reliability and security of the entire system, even if individual nodes are attacked. The cryptographic features and decentralized nature of blockchain make it an ideal solution for authentication and access to services in the IoMT context. Blockchain can be categorized into three main types: public, private, and consortium.
A public blockchain allows all stakeholders, including miners and users, to access blocks and transactions. However, in healthcare, patient privacy must be safeguarded, making a public blockchain less suitable. On the other hand, a private blockchain is permissioned, meaning only registered participants can join the network. It imposes more restrictions and offers scalability, requiring prior permission for network entry. In other words, membership in the private blockchain necessitates authorization from the controlling organization, group, or individual. The network administrator defines the level of access, authority, and operational guidelines within this network.
In the proposed protocol, a private blockchain model is employed. The user's smart gateway/mobile terminal stores the smart card on the blockchain and maintains the encrypted block address, enabling access to the user's smart card address through it. This approach ensures the security and integrity of the user's information within the private blockchain network.

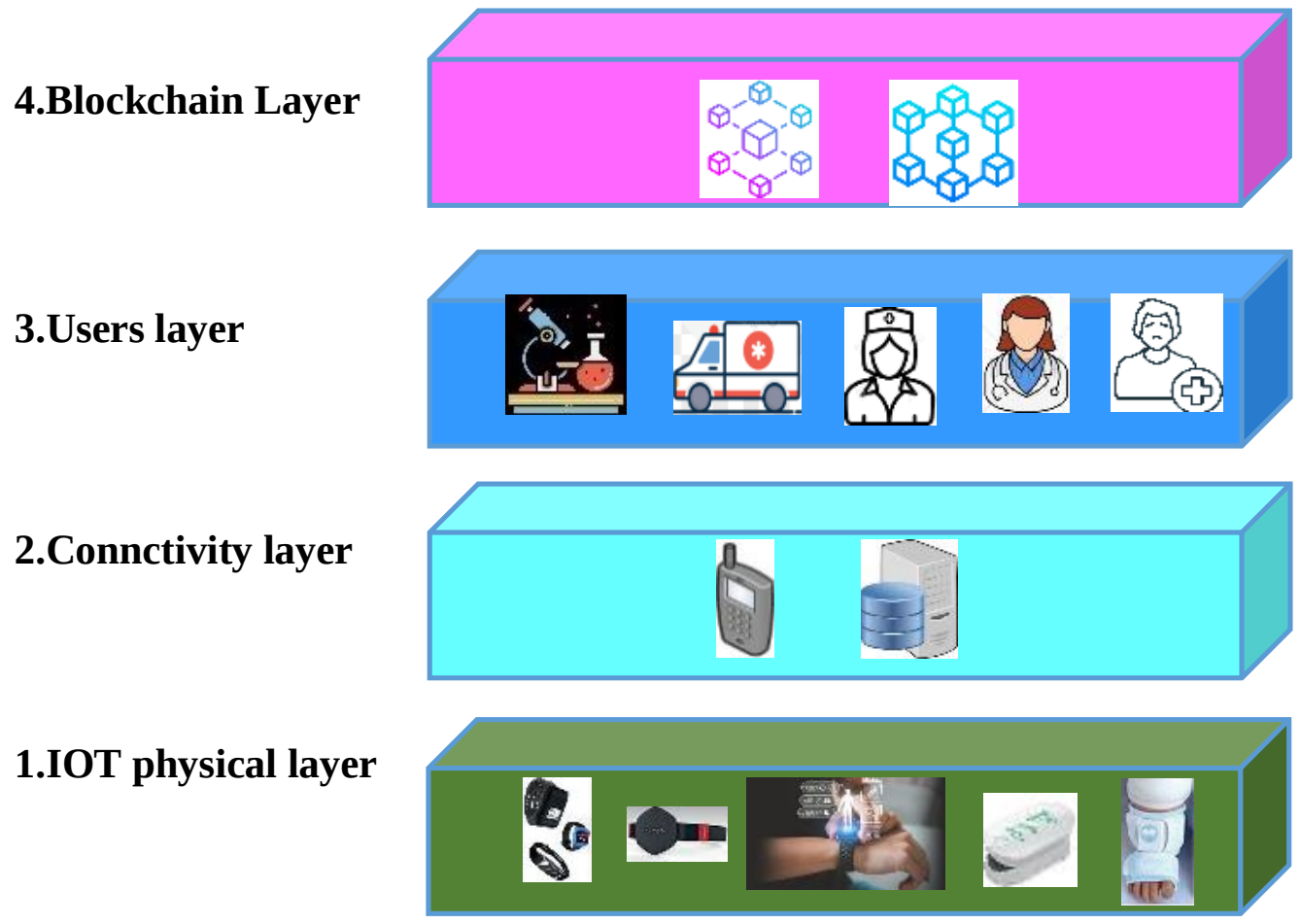

**Figure3: Layered architecture of L2AI**

## 2.3. Threat model

In this section, we examine the threat model proposed by Dolev and Yao [23], which is widely employed in existing schemes. This model encompasses various aspects of network security and considers potential malicious activities by attackers. The key elements of this threat model are as follows:

- IoT users and sensors are considered unreliable within this model.
- The communication channel and interactions between parties are assumed to be insecure from the outset.
- A malicious attacker has the capability to eavesdrop on all communication messages, provide false replies, and modify or delete intercepted messages.
- The attacker can discover the password or parameters of a smart card, although not simultaneously.
- Attacks such as impersonation and man-in-the-middle attacks can be executed by a malicious attacker.

- The attacker can physically capture deployed IoT sensors and gain access to all sensitive information stored within them.
- When three user security factors are verified, the attacker can obtain two of these factors.
- It is assumed that system management and servers are reliable and secure.

By considering these factors, the Dolev and Yao threat model provides insights into the potential vulnerabilities and security challenges present in the network.

### 2.4. Authorization mechanism

To safeguard sensitive information and system resources, various authorization mechanisms are employed as access control solutions, ensuring that only authorized users can access specific resources. Within the context of hospital and patient access policies, system management defines the permitted actions for users regarding IoMT data. In the L2AI, we utilize a lightweight authorization strategy to achieve the desired access control. Our mechanism does not impact system resources and does not require additional communication, providing a seamless and efficient solution.

This access control mechanism ensures that only authorized users can access designated data, which is a fundamental security feature in modern medical applications. In the setup phase of the L2AI, prior to registration, users initiate a request to join the healthcare network through the server. Upon entering the healthcare application and providing their national code, along with selecting the user group, the user receives a token (TG). The System Administration (SA) assigns appropriate ZUi authorization parameters to each user in the TG format. The ZUi contains information about authorized users and their access privileges to specific data. Additionally, it specifies the patients' data that the user is permitted to access through the server. SA securely stores the TG in the blockchain.

In the L2AI, the server assumes a central role in the access control update mechanism. It can efficiently and promptly update authorization information by transmitting new access control data, granting or revoking access permissions. When the user initiates a login request, the server validates the request and retrieves the authorization parameters stored in the TG corresponding to the user. To make an access decision, the server examines the content of the TG, specifically the ZUi, to verify the user's authorization for accessing the requested real-time data.

The employed access control mechanism is highly suitable for resource-constrained IoT systems, ensuring effective access control while considering the limitations of such systems.

## 3. Preliminaries

### 3.1. Notations

The notations used through this paper are described in Table 2.

**Table 2 Notations used in this paper.**

| Symbol | Explanation |
|---|---|
| $U_i$ | $i$th user |
| HMS | Hospital Server |
| $S$A | System administration |
| $T_{Gi}$ | $Ui$ 's authorization parameter |
| $ID_{HMS}$ | HMS unique identity |
| $S_{HMS}$ | HMS symmetric encryption key |
| $ID_i$ , $PW_i$ , $BIO_i$ | $Ui$ 's identity, password and biometric information, respectively |
| $D_{TIDi}$ | Pseudo identity of $U_i$ |
| $AX_{U_i}$ | 160−bit dynamic index of $U_i$ |
| $ZU_i$ | $Ui$ 's authorization parameter |
| $T1$ , $T2$ | Various current timestamps |
| $\Delta T$ | Maximum transmission delay |
| $R^1_{HMS}$ , $R^2_{HMS}$ | Random secret generated by Hospital server |
| $N_C$ , $N_s$ | Nonce generated by Hospital server, Device |
| $sK_{is}$ | Session key between $U_i$ and HMS |
| $Gen(.)$ | Fuzzy extractor generation method |
| $Rep(.)$ | Fuzzy extractor reproduction method |
| $BIO_i$ | $Ui$ 's personal biometrics |
| $\sigma_i$ | $Ui$ 's biometric secret key |
| $\tau_i$ | $Ui$ 's public reproduction parameter |
| $h(.)$ | Cryptographic collision-resistant one-way hash function |
| $\|$ | Concatenation operation |
| $\oplus$ | Bitwise XOR operation |
| Enc (.) | Symmetric encryption function |
| Dec (.) | Symmetric decryption function |
| SC/BC | Smart card / Blockchain |

### 3.2. Hash functions

A cryptographic one-way hash function [24] is defined as: $h : X \to$ , where $X = \{0, 1\}*$ , and $Y = \{0, 1\}n$ . The function $h$ maps an arbitrary length string of bits (denoted by $X$, as input message) to a fixed length string (denoted by , as output message) called the hashed value of size $n$. Hash functions are key elements in many cryptographic applications especially in authentication schemes. The hash function satisfies the following properties:

1. Easiness: Given $m \in X$, it is easy to compute $y$ such that $y = h(m)$.

2. Preimage resistant: It is hard to find $m$ from given $y$, where $h(m) = y$.

3. Collision resistant: It is hard to find a pair $(m, m')$ ∈ $X$ such that $h(m) = h(m')$, where $m \neq m'$ .

4. Mixing-transformation: With any input $m \in X$, the hashed value $y = h(m)$ is computationally indistinguishable from an uniform binary string in the interval $\{0, 2\ n\}$, where $n$ is the output length of hash $h(.)$.

### 3.3. Fuzzy extractor

As a multi-factor authentication scheme, in (L2AI), we use the widely accepted fuzzy extractor technique [25] to verify the user's identity.

The fuzzy extractor consists of two algorithms:

1. Probabilistic generation algorithm ($Gen$) to extract uniformly random bits ( ) and public information ($\tau i$ ) from the biometric template ($BIOi$ ) as original user's biometric data obtained at the registration process. $Gen(BIOi) = (\sigma i , \tau i)$.

2. Deterministic reproduction algorithm ($Rep$) to recover the original biometric key data ($\sigma i$ ) from a noisy biometric ($BIO*\ i$ ) as current user's biometric data obtained at the login process corresponding to the computed public information ($\tau i$ ) and predefined error tolerance between original and current user's biometric information. $\sigma i = Rep(BIO*\ i , \tau i)$.

### 4. The proposed scheme L2AI

In this section, we present a lightweight multi-factor authentication and authorization scheme for real-time data access on a blockchain platform called L2AI. The protocol utilizes only hash functions and bitwise

XOR operations. The three agents involved in L2AI are as follows: (1) the user's smart card, denoted as $SC_{Ui}$, (2) the user's password, denoted as $PW_i$, and (3) the user's personal biometrics, denoted as $BIO_i$.
The protocol consists of four main stages: the registration stage, the login and authentication stage, the biometric and password update stage, and the access level update stage. Each of these stages is described in detail in the following subsections. The symbols used in the proposed protocol are presented in Table 2.

**4.1. Setup phase**

During the setup phase, the system management of the SA selects a unique identity identifier ($ID_{HMS}$) and generates a 160-bit secret key ($S_{HMS}$) for private encryption for each hospital server before their deployment.

**4.2. User registration phase**

In order to enhance security during the communication process between users and hospital servers, users utilize a smart gateway or a mobile phone terminal to register with the hospital server, as depicted in Figure 1. When a new user, Ui, intends to access real-time data legitimately through the relevant server on the blockchain platform, they initiate the registration process by sending a request message to the server via the application. This message includes user information such as phone number, Ui credentials (identification card) for identity verification with SA, and the selection of the user group. Additionally, the message requests a token (user license) from the server.
The SA system management determines the user's permissions and privileges based on the required access level and Table 3, and subsequently sends this information to the user. Each TGi token is a unique 160-bit dynamic index. With the token in hand, the user can proceed to register on the hospital's servers. As part of the registration process, the server issues a digital smart card to the new user, enabling them to utilize the services provided. It is important to note that the communication channels during the registration phase are insecure. The registration step in our protocol consists of four distinct steps, which are summarized in Figure 4.

**• Step Reg1.**

After receiving the user's initial request, the server generates the $T_{Gi}$ token for the user. Subsequently, the server calculates the hash of $T_{Gi}$ (as per equation 1) and encrypts it using its private key ($S_{HMS}$) following Equation 2. The encrypted token and its hash are then stored in the blockchain. Finally, the server sends an SMS containing the $T_{Gi}$ to the user's mobile number.

$$X=h(T_{Gi}) \quad (1)$$

$$y=Enc_{SHMS}(T_{Gi}) \quad (2)$$

**• Step Reg2.**

After receiving $T_{Gi}$, the user proceeds to choose their username ($ID_i$), password ($PW_i$), and biometrics ($B_i$). The user then calculates the hash of the biometrics ($B_i$) and the hash of $T_{Gi}$ using Equations 3 and 4, respectively. Additionally, the user calculates the hash function of the hidden password $PWD_i$, as described in equation (5) below. To protect the username from potential attacks, the user encloses $ID_i$ within $DID_i$, following Equation 6.

$$b_i=H(B_i) \quad (3)$$

$$X=h(T_{Gi}) \quad (4)$$

$PWD_i=h(PW_i \| b_i)$ (5)

$DID_i=ID_i \oplus h(x \| T_{Gi})$ (6)

Finally, the user's smart gateway/mobile terminal sends the message X, $DID_i$, $PWD_i$ to the server via the public channel.

**•Step Reg3.**

As soon as the server receives the information sent by the user, it checks if Any(X) is established in the blockchain. If it is not, the registration attempt is immediately terminated. Otherwise, the server retrieves the token ($T_{Gi}$) using equation 7 and obtains the user $ID_i$ from equation 8.

Next, the server generates $R^1_{HMS}$ (Random Number) and conceals the $D_{TIDi}$ (pseudo-identity of the user) and the user's access level within the AXui parameter, following Equation 10. The server also calculates the $EID_i$, $K_i$, and $HID_{HMS}$ parameters. Subsequently, a smart card {$K_i$, $EID_i$, h(.), $HID_{HMS}$, $R^1_{HMS}$, $AX_{ui}$} ($SC_{Ui}$) is issued to the user and sent through the public channel. To prevent frequent registration of the user's identity in subsequent stages of use, $h(D_{TIDi})$ and $ID_i$ are stored in the blockchain.

$T_{Gi}=Dec_{HMS}(y)$ (7)

$ID_i=DID_i \oplus h(x \| T_{Gi})$ (8)

$D_{TIDi}=ID_i \oplus R^1_{HMS}$ (9)

$AX_{ui}= T_{Gi} \oplus (D_{TIDi} \| ID_{HMS})$ (10)

$K_i=h(S_{HMS} \| ID_i) \oplus PWD_i$ (11)

$EID_i= D_{TIDi} \oplus h(S_{HMS})$ (12)

$HID_{HMS} =h(ID_{HMS} \| S_{HMS}) \oplus D_{TIDi}$ (13)

**• Step Reg4.**

After receiving the smart card from the server, the user proceeds to obtain their $D_{TIDi}$ (pseudo-identity). Subsequently, the user performs the necessary calculations to generate the parameters required for authentication in the next step. Finally, the parameters {$E_i$, $F_i$, $EID_i$, $R^1_{HMS}$, $HID_{HMS}$, $AX_{ui}$} are added to the smart card and stored in the blockchain. Additionally, the encrypted address of the desired block is returned and stored in the smart gateway/mobile terminal.

$D_{TIDi}=ID_i \oplus R^1_{HMS}$ (14)

$F_i=h(PWD_i \oplus K_i \oplus b_i)$ (15)

$F_i=h(PWD_i \oplus K_i \oplus b_i)$ (16)

Table 3. Access permission

| ROLE | Authorization GROUP | authorization parameter |
|---|---|---|
| USER_Doctor | D | $T_G = T_D$ |
| USER_Nurse | N | $T_G = T_N$ |
| USER_Patient | P | $T_G = T_P$ |
| USER_DRAG | M | $T_G = T_M$ |
| USER_Hospital | H | $T_G = T_H$ |
| USER_SA | SA | $T_G = T_{SA}$ |
| USER_Emergency | E | $T_G = T_E$ |
| USER_Laboratory | L | $T_G = T_L$ |

| **Hospital server** | | **user** |
|---|---|---|
| Create unique $T_{Gi}$<br>Select $ID_{HMS}$,$S_{HMS}$ اسم و کلید خصوصی بیمارستان<br>Computes<br>$y=Enc_{SHMS}(T_{Gi})$<br>$X=h(T_{Gi})$<br>Store (x, y) in BC | $T_{Gi}$ Send By Mobile Number → | |
| | ← $\{X, DID_i, PWD_i\}$ | Receive $T_{Gi}$<br>Select <$ID_i$,$PW_i$,$B_i$><br>Computes<br>$b_i=H(B_i)$<br>$X=h(T_{Gi})$<br>$PWD_i=h(PW_i \Vert b_i)$<br>$DID_i=ID_i \oplus h(x \Vert T_{Gi})$ |
| If (! Any(X)) in BC<br>Reject<br>Else<br>Computes<br>$T_{Gi}=Dec_{HMS}(y)$<br>$ID_i=DID_i \oplus h(x \Vert T_{Gi})$<br>Generate $R^1_{HMS}$<br>Computes<br>$D_{TIDi}=ID_i \oplus R^1_{HMS}$<br>$AX_{ui}= T_{Gi} \oplus (D_{TIDi} \Vert ID_{HMS})$<br>$K_i=h(S_{HMS} \Vert ID_i) \oplus PWD_i$<br>$EID_i= D_{TIDi} \oplus h(S_{HMS})$<br>$HID_{HMS} =h(ID_{HMS} \Vert S_{HMS}) \oplus D_{TIDi}$<br>Create $SC_{Ui}=\{ K_{i,}, EID_i, h(.), HID_{HMS}, R^1_{HMS.}, AX_{ui} \}$<br>Store($h(D_{TIDi})$,$ID_i$) in BC<br>Send Smart Card to User | $SC_{Ui}$ → | |
| | | $D_{TIDi}=ID_i \oplus R^1_{HMS}$<br>$F_i=h(PWD_i \oplus K_i \oplus b_i)$<br>$E_i=K_i \oplus h(PWD_i \Vert b_i)$<br>$SC_{Ui}=\{E_i, F_i, EID_i, R^1_{HMS}, HID_{HMS}, AX_{ui} \}$ |

**Fig. 4. Summary of user registration phase.**

## 4.3. User Log-in phase

A legitimate user, $U_i$, who possesses a valid $SC_{Ui}$, must prove their legitimacy to the smart gateway/mobile terminal in order to access IoMT services and log into system resources. To accomplish this, the following steps should be followed, as summarized in Figure 5.

**• Step Log1.**

Initially, the user enters their identity $ID_i$, password PWi, and biometric $B_i$. The smart gateway/mobile terminal retrieves the user's smart card information, $SC_{Ui}$, from the encrypted address of the desired block, with the assistance of the blockchain management intermediary. Subsequently, the biometric hash $b^*_i$, secret code hash $PWD_i^*{}_i$, and pseudo-identity $D^*_{TIDi}$ are calculated using equations (19)-(17). The parameters $F^*_i$ and $K_i$* are then computed to validate the legitimacy of the user's information, as per equations (21)-(20).

$$b^*_i=H(B_I) \quad (17)$$

$$PWD_i^*=h(Pw_i \| b^*_i) \quad (18)$$

$$D^*_{TIDi}=ID^*_i \oplus R^1_{HMS} \quad (19)$$

$$K_i^*=E_i \oplus h(PWD*_i \| b^*_i) \quad (20)$$

$$F^*_i=h(PWD^*_i \oplus K^*_i \oplus b^*_i) \quad (21)$$

The smart gateway/mobile terminal verifies the equality of the $F^*_i$ value with the stored $F_i$ value in the smart card to determine if the user is the rightful owner of $SC_{Ui}$. If these two values are not equal, the login attempt is immediately terminated. Otherwise, the user's identity, $U_i$, is confirmed.

**• Step Log2.**

User $U_i$ generates a timestamp $T_1$ and calculates the following parameters.

$$C_i=K^*_i \oplus PWD^*_i \quad (22)$$

$$h(ID_{HMS} \| S_{HMS}) = HID_{HMS} \oplus D^*_{TIDi} \quad (23)$$

$$W_1=h(D^*_{TIDi} \|(h(ID_{HMS} \| S_{HMS}))) \quad (24)$$

$$M_1=h(C_i \| T_1 \| W_1) \quad (25)$$

Finally, the smart gateway/mobile terminal sends the login request message $MSG_1 = \{T_1, M_1, EID_i, AX_{ui}\}$ through the public channel to the server.

### 4.4. Authentication and key exchange phase

In L2AI, we have designed an efficient authentication process to establish authentication between hospital servers and IoMT users. The purpose of this step is to ensure secure communication and data encryption between the parties involved. At the end of a successful authentication and key exchange phase, a session key is generated between the corresponding hospital server and the user $U_i$. This session key enables real-time communication with enhanced security through the public channel. Figure 5 provides a summary of the authentication and key exchange stage.

**• Step Auk1.**

Upon receiving the login request message MSG1 = {T1, M1, EIDi, AXui} from user Ui, the server first checks the value of T1 using the condition $(|T_S-T_1|)>\Delta T$ in equation (26). This condition ensures that the message is new, where $T_1$ represents the time of message reception and $\Delta T$ is the predetermined maximum transmission delay. If the condition is not satisfied, the session is immediately terminated.

$$\text{If}(|T_S-T_1|)>\Delta T \quad (26)$$

Otherwise, the server calculates the pseudo-identity and the access token $T_{Gi}$ of user $U_i$ using equations (27) and (28) respectively. The server then verifies the validity of the hash of both parameters by checking the blockchain. If there is no match, the session is terminated. However, if the hash values are valid, the server further checks the user's access permission for the activity scope and time of the access level specified by the $T_{Gi}$ token. If the access level permission is approved, the server retrieves the user's identity $ID_i$ from the blockchain based on the hash of the pseudo-identity.

$$D^{**}_{TIDi}= EID_i \oplus h(S_{HMS}) \quad (27)$$

$$T_{Gi}= AX_{ui} \oplus (D^{**}_{TIDi} \| ID_{HMS}) \quad (28)$$

Next, to authenticate the user, the server calculates the parameters $C_i$ and $W^*_1$ using equations (29) and (30) respectively, followed by computing $M^*_1$ using equation (31).

$$C_i=h(S_{hms} \| ID_i) \quad (29)$$

$$W^*_1=h(D^{**}_{TIDi} \| (h(ID_{HMS} \| S_{HMS}))) \quad (30)$$

$$M^*_1=h(C^*_i \| T_1 \| W^*_1) \quad (31)$$

If the condition $M^*_1 = M_1$is not satisfied, the connection is terminated. However, if $M^*_1$ and $M_1$ are equal, the server ensures that the message MSG1 = {T1, M1, EIDi, AXui} sent by the user has not been tampered with. In other words, the conditions $M^*_1 = M_1$and the authorization check for $D^{**}_{TIDi}$: $T_{Gi}$ are met, confirming the user's authentication and authorization by the IoMT server. The server then generates a random number Ns and a timestamp T2, and calculates the parameters $M_2$, $M_3$, and the session key as follows:

$$SK=h(W^*_1 \| N_S) \quad (32)$$

$$M_2= SK \oplus W^*_1 \quad (33)$$

$M_3=h(C_i \| T_2 \| W^*_1 \| SK)$ (34)

In order to enhance user anonymity, L2AI generates a new pseudo-identity for the user in each session by using a new random number value. The new parameters are calculated as follows:

$D^{new}_{TIDi}=ID_i \oplus R^2_{SHM}$ (35)

$AX^{new}_{ui=} T_G \oplus (D^{new}_{TIDi} \| ID_H$ (36)

$EID^{new}_i= D^{new}_{TIDi} \oplus h(S_{HMS}$ (37)

$HID^{new}_{HMS} =h(ID_{HMS} \| S_{HMS}) \oplus D^{new}_{TIDi}$ (38)

Subsequently, the server updates the SCUi smart card by replacing the previous parameters with the new ones, including $R^2_{SHM}$, $EID^{new}_i$, $AX^{new}_{ui}$, and $HID^{new}_{HMS}$. Additionally, the new pseudo-identity hash h(DnewTIDi) replaces the previous one. Finally, the server sends the message $\{M_3, M_2, T_2\}$ to user Ui through an insecure channel.

**• Step Auk2.**

After receiving the message $\{M_3, M_2, T_2\}$ (referred to as $MSG_2$), user Ui checks $T_2$ by applying the condition ($|T_C - T_1| > \Delta T$) stated in equation (39). Here, $T_2$ represents the time of receiving the message $\{M_3, M_2, T_2\} = MSG_2$, and $\Delta T$ denotes the predetermined maximum transmission delay. If the condition is not satisfied, the session is terminated immediately. However, if the condition is met, Ui proceeds to calculate the shared session key using Equation (40). Subsequently, it computes $M^*_3$ based on equation (41) and checks whether $M^*_3$ is equal to $M_3$. If the conditions are not met, the session is terminated immediately. On the other hand, if the conditions are fulfilled, the server is authenticated by user Ui, and the shared session key is verified.

$If(IT_C-T_1I)>\Delta T$ (39)

$SK=M_2 \oplus W_1$ (40)

$M^*_3=h(C_i \| T_2 \| W_1 \| SK)$ (41)

| **user** | | **Hospital server** |
|---|---|---|
| User $U_i$ input $ID_i, Pw_i, B_i$<br>Device Get Smart Card by hash address<br>Computes:<br>$b^*_i = H(B_I)$<br>$PWD_i^* = h(Pw_i \parallel b^*_i)$<br>$D^*_{TIDi} = ID^*_i \oplus R^1_{HMS}$<br>$K_i^* = E_i \oplus h(PWD^*_i \parallel b^*_i)$<br>$F^*_i = h(PWD^*_i \oplus K^*_i \oplus b^*_i)$<br>if $(F^*_i \not\equiv F_i)$ Reject<br>Else<br>Generate a Time stamp $T_1$<br>Compute<br>$C_i = K^*_i \oplus PWD^*_i$<br>$h(ID_{HMS} \parallel S_{HMS}) = HID_{HMS} \oplus D^*_{TIDi}$<br>$W_1 = h(D^*_{TIDi} \parallel (h(ID_{HMS} \parallel S_{HMS})))$<br>$M_1 = h(C_i \parallel T_1 \parallel W_1)$ | $MSG_1 = \{T_1, M_1, EID_i, AX_{ui}\}$<br>→ | |
| If$(IT_C - T_1I) > \Delta T$<br>Reject<br>Else<br>Compute<br>$SK = M_2 \oplus W_1$<br>$M^*_3 = h(C_i \parallel T_2 \parallel W_1 \parallel SK)$<br>If $(M^*_3 \not\equiv M_3)$<br>Reject<br>Else,Session key is created correctly | $MSG2 = \{M3, M_2, T_2\}$<br>← | If$(IT_S - T_1I) > \Delta T$ reject<br>Else<br>Compute<br>$D^{**}_{TIDi} = EID_i \oplus h(S_{HMS})$<br>$T_{Gi} = AX_{ui} \oplus (D^{**}_{TIDi} \parallel ID_{HMS})$<br>(If (! Any(h($D^{**}_{TID}$)i& h($T_{Gi}$))) Reject<br>Else<br>Checks authorization $D^{**}_{TIDi}$: $T_{Gi}$<br>If authorized, Compute<br>Get $ID_i$ By h( $D_{TIDi}$)<br>$C_i = h(S_{hms} \parallel ID_i)$<br>$W^*_1 = h(D^{**}_{TIDi} \parallel (h(ID_{HMS} \parallel S_{HMS})))$<br>$M^*_1 = h(C^*_i \parallel T_1 \parallel W^*_1)$<br>If $(M^*_1 \not\equiv M_1)$<br>Reject<br>Else<br>Generate a random number $N_C$,Time stamp $T_2$<br>Compute<br>$SK = h(W^*_1 \parallel N_S)$<br>$M_2 = SK \oplus W^*_1$<br>$M_3 = h(C_i \parallel T_2 \parallel W^*_1 \parallel SK)$<br>Generate $R^2_{SHM}$<br>$D^{new}_{TIDi} = ID_i \oplus R^2_{SHM}$<br>$AX^{new}_{ui} = T_G \oplus (D^{new}_{TIDi} \parallel ID_{HMS})$<br>$EID^{new}_i = D^{new}_{TIDi} \oplus h(S_{HMS})$<br>$HID^{new}_{HMS} = h(ID_{HMS} \parallel S_{HMS}) \oplus D^{new}_{TIDi}$<br>Replace $R^2_{SHM}$ , $EID^{new}_i$ , $AX^{new}_{ui}$ , $HID^{new}_{HMS}$ of $SC_{Ui}$<br>Replace $h(D_{TIDi})$ BY $h(D^{new}_{TIDi})$ |

**Fig. 5. Summary of login and authentication and key agreement phases.**

## 4.5. Password and biometric update phase

In L2AI, an authorized user can independently update their password and/or biometric information without the need to communicate with the SA (Security Administrator). The user should follow the steps outlined below to update their password and biometrics without involving the update server. Please refer to Figure 6 for a summary of this process.

• **Step PB1.**

First, the user enters their identity $ID_i$, current password $PW^{old}_i$, and biometrics $B^{old}_i$. The smart gateway/mobile terminal then retrieves this information using the user's smart card $SC_{Ui}$ and performs the necessary calculations based on equations (42) - (45) to obtain the foldi parameter.

$$b^{old}_i=H(B^{old}_I) \quad (42)$$

$$PWD^{old}_i=h(PW^{old}_i \| b^{old*}_i) \quad (43)$$

$$K^{old}_i=E_i \oplus h(PWD^{old}_i \| b^{old}_i) \quad (44)$$

$$F^{old}_i=h(PWD^{old}_i \oplus K^{old}_i \oplus b^{old}_i) \quad (45)$$

Next, it checks the equality of $F^{old}_i$ and $F_i$. If these values are not equal, the password and biometric update phase is immediately terminated. Otherwise, $U_i$ is recognized as valid and ready to update its secret parameters.

**• Step PB2**

Step 2: Now, the smart gateway/mobile terminal of user $U_i$ requests the new password $PW^{new}_i$ and new biometric Bnewi from $U_i$. $U_i$ selects a new password PWnewi and a new biometric $B^{new}_i$. Finally, the device calculates the new parameters based on equations (46) - (50). These new parameters replace the old ones ($K^{old}_i$, $Fold_i$, $E^{old}_i$, $PWD^{old}_i$, $b^{old}_i$) in the smart card $SC_{Ui}$.

$$b^{new}_i=H(B^{new}_i) \quad (46)$$

$$PWD^{new}_i=h(PW^{new}_i \| b^{new}_i) \quad (47)$$

$$K^{new}_i= K^{od}_i \oplus PWD^{old}_i \oplus PWD^{new}_i \quad (48)$$

$$F^{new}_i=h(PWD^{new}_i \oplus K^{new}_i \oplus b^{new}_i) \quad (49)$$

$$E^{new}_i=K^{new}_i \oplus h(PWD^{new}_i \| b^{new}_i) \quad (50)$$

| **User** | **$Device_{ui}$** |
| --- | --- |
| User $U_i$ input $ID_i$,$PW^{old}_i$,$B^{old}_i$ | Device Get Smart Card by hash address<br>Computes:<br>$b^{old}_i=H(B^{old}_I)$<br>$PWD^{old}_i=h(PW^{old}_i \| b^{old*}_i)$<br>$K^{old}_i=E_i \oplus h(PWD^{old}_i \| b^{old}_i)$<br>$F^{old}_i=h(PWD^{old}_i \oplus K^{old}_i \oplus b^{old}_i)$<br>if ($F^{old}_i \neq F_i$) Reject<br>Else<br>Requst new $PW^{new}_i$and $BIO^{new}_i$ |
| $Ui$ chooses $PW^{new}_i$and $BIO^{new}_i$ | Computes<br>$b^{new}_i=H(B^{new}_i)$<br>$PWD^{new}_i=h(PW^{new}_i \| b^{new}_i)$<br>$K^{new}_i= K^{od}_i \oplus PWD^{old}_i \oplus PWD^{new}_i$<br>$F^{new}_i=h(PWD^{new}_i \oplus K^{new}_i \oplus b^{new}_i)$<br>$E^{new}_i=K^{new}_i \oplus h(PWD^{new}_i \| b^{new}_i)$<br>Replace with {$E^{new}_i$,$F^{new}_i$ } into SC |

**Fig. 6. Summary of password and biometric update phase.**

### 4.6. Authorization update phase

To ensure the security of sensitive system resources, we employ an efficient authorization mechanism for managing system access control. In L2AI, the SA (System Administrator) effectively manages user access control by granting and/or revoking access permissions to registered users. Figure 7 provides a summary of the authorization update step. The following steps provide a detailed description of this process.

**• Step AU1.**

To update the access control for user $U_i$, the SA selects $U_i$'s corresponding $ID_i$ and $h(D_{TIDi})$ from the blockchain. Then, the SA chooses a new token TGi and calculates equations (51) and (52) to update the parameters X and AXui.

$$X^{New}=h(T^{New}_{Gi}) \quad (51)$$

$$AX^{New}_{ui}= T^{New}_{Gi} \oplus (D_{TIDi} \| ID_{HMS}) \quad (52)$$

Subsequently, in the smart card $SC_{Ui}$, the previous value of the $AX^{New}_{ui}$ parameter is replaced, and $X^{New}$ is stored in the blockchain.

**Hospital Server**
Select $ID_i$ , $h(D_{TIDi})$
Generate $T^{New}_{Gi}$
Computes
$X^{New}=h(T^{New}_{Gi})$
$AX^{New}_{ui}= T^{New}_{Gi} \oplus (D_{TIDi} \| ID_{HMS})$
Replace with { $AX^{New}_{ui}$ } into $SC_{Ui}$ , ($X^{New}$) in BC

**Fig. 7. Summary of authorization update phase**